\documentclass[preprint,proceedings]{rmaa}

\SetYear{2006}
\SetConfTitle{Massive Stars: Fundamental Parameters and Circumstellar Interactions}

\title{The Galactic O Star Catalog v.2.0}

\author{
  Alfredo Sota\altaffilmark{1,2}, Jes\'{u}s Ma\'{i}z Apell\'{a}niz\altaffilmark{2}, Nolan R. Walborn\altaffilmark{1}, and Raquel Y. Shida\altaffilmark{3}}

\altaffiltext{1}{Space Telescope Science Institute.}
\altaffiltext{2}{Instituto de Astrof\'{i}sica de Andaluc\'{i}a.}
\altaffiltext{3}{Space Telescope - European Coordinating Facility.}

\suppressfulladdresses

\listofauthors{A. Sota; J. Ma\'{i}z-Apell\'{a}niz; N. R. Walborn; R. Y. Shida}
\indexauthor{Sota, A; Ma\'{i}z-Apell\'{a}niz, J.; Walborn, N. R.; Shida, R. Y. }

\addkeyword{Catalogs}
\addkeyword{Stars: Early-Type} 
\addkeyword{ Stars: Fundamental Parameters}
\begin{document}
\maketitle 

\boldabstract{\ \ \ \ The advent of internet search engines and databases is producing a revolution in astronomy. However, when one compares data from widely diverse sources, the possibility exists that good quality results and poorer ones are being mixed. One of the fields where this danger is more apparent is spectral classification, a subject where lower quality data or the technique of the observer can easily introduce systematic and random errors. }
\vspace{0.3cm}

	A number of lists and catalogs of spectral types for O stars are available in the literature. The works of Morgan et al. (1955), Hiltner (1956), Lesh (1968) , Hiltner et al. (1969), and Garrison et al. (1977) provided original accurate spectral classifications. Those catalogs, though, are far from complete and most of their spectral classifications were done before the introduction of spectral types O3 (Walborn 1971) and O2 (Walborn et al. 2002). Our approach is to build a catalog of O stars with the following conditions. (1) Stars should be selected on the basis of their optical spectroscopy alone. (2) Sources for spectral classification should be accurate and as uniform as possible. (3) Valuable additional information (positions, photometry,  etc.) should be added only if they come from uniform catalogs or if their accuracy can be checked. (4) The catalog should be made accessible through the Web. 

\vspace{0.3cm}

\textbf{Past: v1.0}

	The first version of the GOS catalog (Ma\'{i}z Apell\'{a}niz et al. 2004) was built starting with the 350 Galactic O stars catalogued by one of the authors (N.R.W.) over the years but other known O stars with $V < 8$ were also included, yielding a total of 378 stars. The final list was subdivided in two groups: the main catalog contains the 370 stars which do not have unresolved WR companions while the WR supplement adds 8 stars which are members of such WR+O systems.  
\vspace{0.4cm}

\textbf{Present: v2.0}

	In this new version, we include three new lists. The first one (O-star supplement) adds more than 700 stars that sometimes have been classified as O stars. The second one (WR supplement 2) includes the O stars in O+WR unresolved systems that are not included in the first version of the catalog. Finally a third list (excluded supplement) includes the stars that sometimes have been classified as O stars but have been rejected from the main catalog because they are early B stars. 
\vspace{0.3cm}

\textbf{Immediate future: v2.1+}

	We shall review the new supplement sources for reliability in our judgement, sorting them into three groups on that basis in order to confirm the published classification and add the stars to the main catalog (if the source is reliable), include them in a new list of stars for which we have reservations but we cannot exclude that they are O stars, or transfer them to the excluded supplement (if we confirm that they are not O stars). Other planned tasks are the following:
\\
$\bullet$	 Obtain new digital spectrograms for all the stars in the main catalog and WR supplement 1 (many original spectrograms were photographic) in order to build a spectroscopic atlas.\\
$\bullet$	 Apply the new Hipparcos recalibration for the small subset of stars that are sufficienttly nearby to produce more accurate distances.\\
$\bullet$	 Create queries to search by e.g. magnitude or position.\\
$\bullet$	 Include the catalog in the virtual observatory (VO).\\

	The catalog can be found at: \\
http://www-int.stsci.edu/$\sim$jmaiz/GOSmain.html

\end{document}